# Open-shell TADF: Quartet-derived luminescence with dark radicals

Sebastian Gorgon[1,2,*], Petri Murto[1,3,4], Daniel G. Congrave[3,5], Lujo Matasovic[1], Andrew D. Bond[3], William K. Myers[2], Hugo Bronstein[3], Richard H. Friend[1]

[1]Cavendish Laboratory, University of Cambridge, JJ Thomson Ave, Cambridge, CB3 0HE, United Kingdom
[2]Centre for Advanced Electron Spin Resonance, Department of Chemistry, University of Oxford, Inorganic Chemistry Laboratory, S Parks Rd, Oxford, OX1 3QR, United Kingdom
[3]Yusuf Hamied Department of Chemistry, University of Cambridge, Lensfield Rd, Cambridge, CB2 1EW, United Kingdom
[4]Department of Chemistry and Materials Science, Aalto University, Kemistintie 1, 02150 Espoo, Finland
[5]Department of Chemistry, University of Oxford, Chemistry Research Laboratory, Oxford, OX1 3TA, United Kingdom

*Corresponding author: Sebastian Gorgon, sg911@cam.ac.uk

**Abstract**

High-spin states in organic molecules offer promising tuneability for quantum technologies. Photogenerated quartet excitons are an extensively studied platform, but their applications are limited by the absence of optical read-out via luminescence. Here we demonstrate a new class of synthetically accessible molecules with quartet-derived luminescence, formed by appending a non-luminescent TEMPO radical to Thermally Activated Delayed Fluorescence (TADF) chromophores previously used in OLEDs. The low singlet-triplet energy gap of the chromophore opens a luminescence channel from radical-triplet coupled states. We establish a set of design rules by tuning the energetics in a series of compounds based on a naphthalimide (NAI) core. We observe generation of quartet states and measure the strength of radical-triplet exchange (0.7 GHz). In DMAC-TEMPO, up to 72% of detected photons emerge after reverse intersystem crossing from the quartet state repopulates the state with singlet character. This design strategy does not rely on a luminescent radical to provide an emission pathway from the high-spin state. The large library of TADF chromophores promises a greater pallet of achievable emission colours.

**Introduction**

Organic molecules are an attractive platform for quantum technologies.[1] They offer a highly modular and deterministic approach for constructing precise quantum objects via chemical synthesis.[2] Since structures can be built exclusively from light atoms, long spin coherence times can persist to higher temperatures.[3] The absence of heavy atoms offers further advantages in biocompatibility, increasing the scope for future quantum sensing demonstrations at physiological conditions.

While controlled preparation and manipulation of electron spins in organics is well established, engineering of efficient optical read-out pathways remains a challenge. Light provides the most facile and non-invasive way to read-out a spin system,[4] which would bring organics closer to practical applications.

Key demonstrations of information transfer, storage and manipulation have been made in excited-state multi-spin organic molecules.[5–8] In particular, three-spin systems with quartet ($S = 3/2$) excited states have been explored as a versatile organic qubit platform.[9,10] To date they were obtained by appending a stable, non-luminescent radical to a simple chromophore (Fig. 1a).[11] These quartet candidates contain chromophore units with a locally excited (LE)-type emissive state. Due to large excited-state singlet-triplet ($S_1$-$T_1$) energy gaps, once a quartet state is formed, it cannot repopulate the initially photogenerated, luminescent state. Thus, optical read-out of the high-spin state via emission is impossible in these structures.

Recently, we have shown an alternative approach to constructing quartet-bearing molecules. By employing a luminescent π-conjugated radical and engineering appropriate energy level resonances, our design reversibly linked the quartet state with a doublet state emitting near 700 nm.[12] However, relying on luminescent radicals may restrict the range of emission wavelength tunability,[13] and limit structural and synthetic diversity.

Stable nitroxide radicals like 2,2,6,6-tetramethyl-1-piperidinyloxy (TEMPO) are widely-studied open-shell systems, as they are robust in chemical synthesis, electrochemical applications and under light-excitation.[14,15] TEMPO has been incorporated into a wide range of chemical structures through mild chemistry that circumvents the need for additional radical conversion steps, making it structurally diverse and synthetically accessible.[10] However, as typical of σ-radicals, TEMPO acts as a luminescence quencher.[16] Existing reports of light emission in TEMPO-derived species rely on harsh redox reactions or aggregation-based strategies.[17,18]

TADF is currently a central direction in organic optoelectronic research.[19,20] This allows us to take advantage of the vast array of TADF structures to unlock a new direction for luminescent high-spin molecules. As established for π-conjugated radicals, reversibly linking the quartet state with an emissive state requires minimising their energy offset. In TADF-TEMPO structures, the small energy offset requirement is automatically fulfilled due to the small $S_1$-$T_1$ gap inherent to TADF materials (Fig. 1b). The additional radical-triplet exchange only calls for an interaction on the order of a few μeV, to ensure that a strongly coupled quartet state is formed. As the emissive channel is provided by the TADF chromophore, we do not require a luminescent radical.

We focus on a highly efficient TADF operating near 2.0 eV with a high intersystem crossing yield, which is a donor-acceptor system with strong charge transfer (CT) character.[21] The design consists of an electron-accepting 1,8-naphthalimide (NAI) core with its nitrogen substituted with either a cyclohexane (Cy) or near-isostructural TEMPO radical as $R_2$ (Fig. 1c). To explore the sensitivity of the mechanism to the energetics, we prepare a series of structures with three different electron-donating units $R_1$ coupled at the 4-position: 3,6-di-*tert*-butylcarbazole (Cz),[22] 9,9-dimethyl-9,10-dihydroacridine (DMAC),[21] and phenoxazine (Phx),[23] in decreasing order of CT energy.

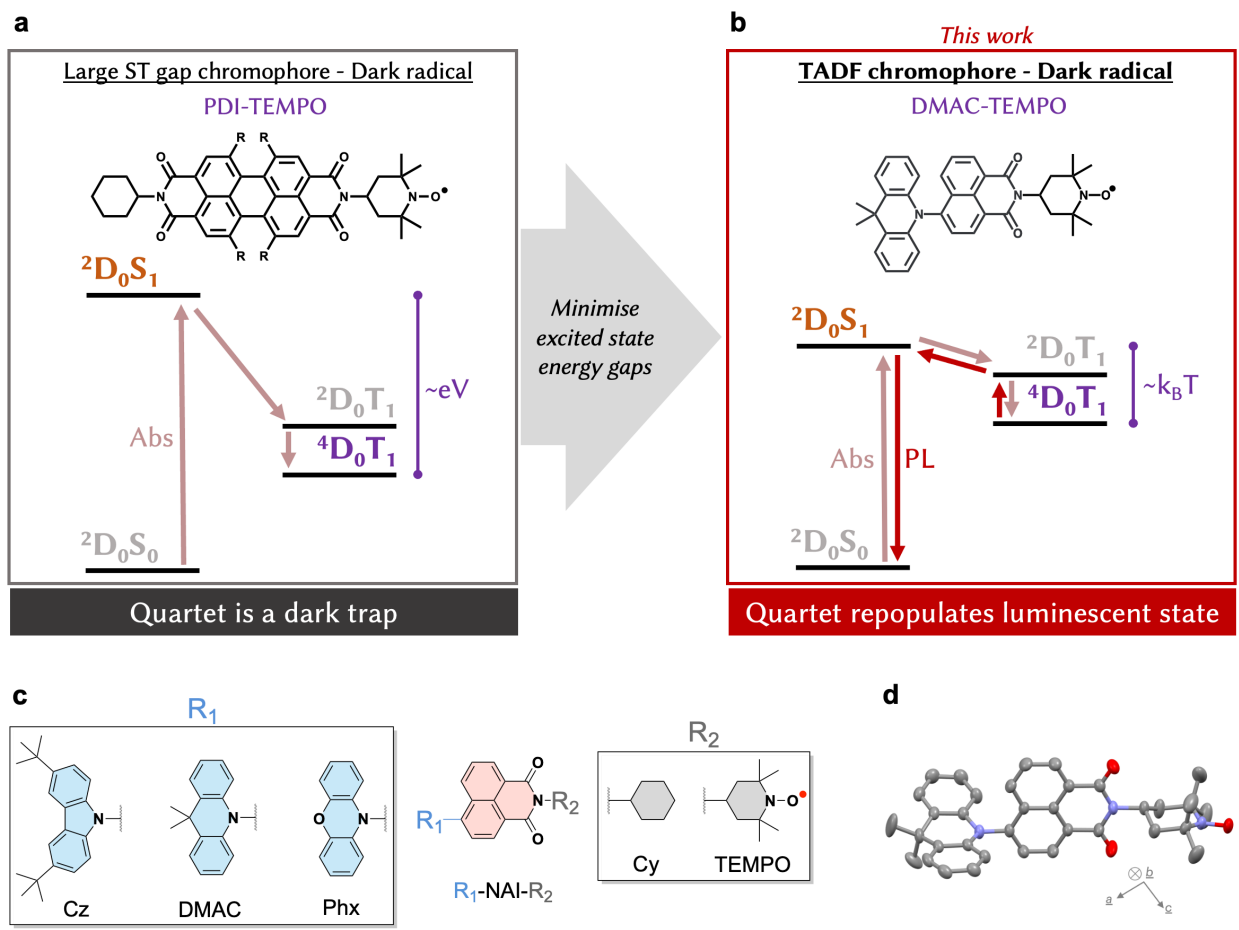

**Fig. 1. Linking high-spin and luminescent states in radicals.** Simplified Jabłoński diagrams for organic molecules with quartet, $^4D_0T_1$, excited states. **a,** Structures based on dark radicals reported to date, such as PDI-TEMPO, have large singlet-triplet (ST) gaps, therefore the luminescent $^2D_0S_1$ is inaccessible from the quartet. **b,** In the new strategy presented here, we instead use a TADF chromophore with a small ST gap. The luminescent $^2D_0S_1$ is thermally accessible from the quartet and delayed emission is observed, despite the radical being non-luminescent. **c,** Molecular structures explored for the TADF-TEMPO strategy consist of an NAI core coupled to $R_1$, an electron donating unit: either Cz-, DMAC-, or Phx-. We study the radical and non-radical (-Cy) compounds, the latter acting as a reference. **d,** Molecular structure of DMAC-TEMPO taken from the X-ray crystal structure, revealing the large dihedral angle between the DMAC and NAI units (82.4°).

**Results**

Both the DMAC-substituted molecules, DMAC-Cy and DMAC-TEMPO, and the Phx-substituted molecules, Phx-Cy and Phx-TEMPO, were synthesized via Buchwald–Hartwig amination using a palladium catalyst and a bulky alkoxide base with yields varying between 60–90%.[24,25] The Cz donor was less reactive in these conditions giving low yields (5% or less). Synthesis of Cz-Cy and Cz-TEMPO required deprotonation of the Cz nitrogen with a Grignard reagent, and these molecules were isolated with yields of 20–30%.[26] NMR spectroscopy suggests that coupling of the radical electron to the NAI hydrogen nuclei is stronger than its coupling to the three donors' hydrogen nuclei, which is observed as significant broadening of the $^1$H NMR signals of the NAI unit rather than those of the donor units (Fig. S1). The NMR data can be rationalised by considering that the radical is directly linked to the acceptor, whereas interactions between the radical electron and the relevant donor nuclei are weakened by the large dihedral, as shown for DMAC-TEMPO in Fig. 1d. Detailed synthetic protocols, X-ray crystal structures and NMR spectra are provided in the Supplementary Information file (Figs. S2-10).

CT-type photoluminescence is observed for all 6 compounds (Fig. 2a), as expected for this TADF motif.[21,27] The emission lineshapes do not significantly evolve as function of time after photoexcitation at room temperature (Fig. S13), suggesting the PL arises from a single excited state in each molecule. For the Cy-derivatives, the emission wavelengths at room temperature in toluene solutions follow expected red-shifts with increasing strength of the donor moiety $R_1$, spanning a range of 0.35 eV. The corresponding TEMPO-derivatives show a very small (ca. 40 meV) redshift. Cyclic voltammograms (Fig. S11) and absorption spectra (Fig. S12) are consistent with these trends in the emission spectra. Substituting TEMPO onto the acceptor results in a <0.1 V shift of its reduction to a less negative potential (deeper LUMO energy), whereas increasing the donor strength pushes the oxidation to a less positive potential (shallower HOMO energy). Hence, substituting TEMPO and increasing the donor strength both have the effect of lowering the HOMO-LUMO energy gap.

DMAC-Cy shows the highest proportion of delayed fluorescence, with over 90% of all emitted photons arriving via delayed emission in 1% in PMMA films (Table 1). Its PLQE of 35% is expected given the low emission energy and is consistent with similar derivatives synthesised to-date.[21,28] DMAC-TEMPO preserves these dynamics (Fig. 2b and S14).

To investigate the energetic landscape around the emissive state, we perform temperature-dependent luminescence spectroscopy. In DMAC-TEMPO, the fast emission component has a lifetime of ca. 13 ns and contributes around 27% of all emitted photons across the entire 10–292 K temperature range (Fig. 2c and S15). This indicates it is due to $^2D_0S_1^{(CT)}$ prompt fluorescence in conformations where this outcompetes quartet state generation. The delayed emission in DMAC-TEMPO is strongly temperature-activated, with delayed emission intensity approximately 5 times larger at room temperature compared to 10 K. The emission lineshape and time-dependence of emission wavelength do not depend on temperature. This confirms that luminescence occurs from the same emissive state at all temperatures. Arrhenius analysis reveals that this $^2D_0S_1^{(CT)}$ state can be reformed after crossing an activation barrier of 39±9 meV, which is approximately $1.5k_BT$ at room temperature. This is similar to the literature values for the activation energy on the related radical-free DMAC-NAI TADF compounds.[29,30]

Turning to Cz-TEMPO, at 10 K we observe a ms lifetime emission which is redshifted by around 0.4 eV from the prompt luminescence and exhibits a pronounced vibronic structure (Fig. S16). We thus assign it to phosphorescence from the NAI $^3$LE-character state, consistent with literature.[27] As this structural motif is common across all 6 compounds, this measurement indicates its energetic position for the whole series. This confirms that the NAI $^3$LE-character state is near-isoenergetic with the CT character states in both DMAC-Cy and DMAC-TEMPO.

We track the exciton populations via ultrafast transient absorption (TA). Upon photoexcitation, the spectra are dominated by photoinduced absorptions (PIA) near 425 and 720 nm, consistent with literature values for NAI anions (Fig. S17).[31] These initially arise from the photogenerated $S_1^{(CT)}$ in the Cy derivatives and $^2D_0S_1^{(CT)}$ in the TEMPO derivatives. However, due to small ST gaps, at later times the same PIAs may also arise from any states with triplet CT character. These then evolve towards a PIA near 480 nm which corresponds to the LE NAI triplet.[32] We observe an acceleration in the CT PIA depopulation when TEMPO is appended, indicating that the presence of the radical is enhancing intersystem crossing. These dynamics are consistent with the observation of weak prompt fluorescence.

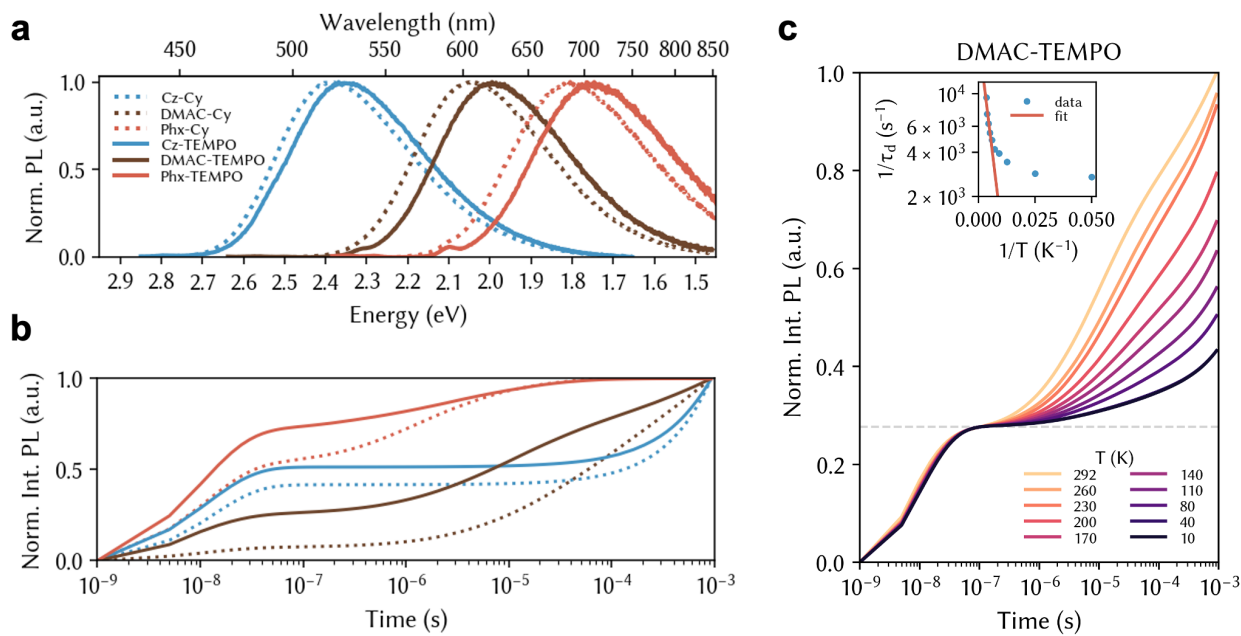

**Fig. 2. Luminescence spectroscopy. a,** Steady-state photoluminescence spectra in 100 µM toluene solutions following near band-edge excitation at 295 K. **b,** Integrated kinetic traces extracted from transient photoluminescence of 1% PMMA films at 293 K under vacuum after 450 nm excitation. **c,** Temperature dependence of emission intensity in 1% DMAC-TEMPO in PMMA film under 400 nm excitation. Luminescence was integrated across the 550-700 nm spectral region. Inset shows Arrhenius fit which extracts an activation energy $E_A$=39±8 meV.

| Material | $E_{PL}$ (eV) | $\lambda_{PL}$ (nm) | $PLQE_{tol}$ (%) | $PLQE_{PMMA}$ (%) | $I_d/I_{tot}$ (%) | $\tau_p$ (s) | $\tau_d$ (s) |
|---|---|---|---|---|---|---|---|
| Cz-Cy | 2.39 | 519 | 22 | 29 | 56 | $1.4\times10^{-8}$ | $1.2\times10^{-2}$ |
| DMAC-Cy | 2.04 | 608 | 35 | 35 | 92 | $1.9\times10^{-8}$ | $1.7\times10^{-5}$ |
| Phx-Cy | 1.81 | 685 | <1 | 4 | 42 | $1.4\times10^{-8}$ | $2.3\times10^{-6}$ |
| Cz-TEMPO | 2.35 | 528 | 1 | 11 | 49 | $1.2\times10^{-8}$ | $3.2\times10^{-3}$ |
| DMAC-TEMPO | 1.99 | 623 | 4 | 10 | 72 | $1.3\times10^{-8}$ | $7.0\times10^{-6}$ |
| Phx-TEMPO | 1.75 | 708 | <1 | 2 | 28 | $1.2\times10^{-8}$ | $4.2\times10^{-6}$ |

**Table 1. Luminescence properties at 292 K.** Emission peak energy ($E_{PL}$) and wavelength ($\lambda_{PL}$), and quantum yield (PLQE) were measured in 100 µM toluene solutions under near-band gap cw excitation. Delayed emission fraction ($I_d/I_{tot}$), as well as prompt ($\tau_p$) and delayed ($\tau_d$) lifetimes were measured in 1% in PMMA films under vacuum under 450 nm pulsed excitation.

Transient ESR (trESR) is a powerful method to study triplet excitons in organic optoelectronic materials.[33,34] The spectra are rich in information on the shape and character of the wavefunction, while the extracted sublevel polarisation pattern can directly determine the type of ISC mechanism.[35]

X-band trESR following 450 nm excitation at 80 K reveals spin-polarised signals in all 6 materials studied here. Starting with the radical-free TADF compounds, Cz-Cy (Fig. 3a) and DMAC-Cy (Fig. 3b) show similar signals with a characteristic polarisation pattern of ISC-generated triplets. Simulation reveals these to have similar dipolar coupling parameters (Table 2). Given the energetics extracted from our luminescence results, we assign this state to the $^3$LE. This matches zfs splittings in the NAI literature.[28,30,32,36] A more complex spectrum is seen for Phx-Cy (Fig. 3c). We can decompose it into a linear combination of the same $^3$LE seen in the other two derivatives, and a second triplet (Fig. S18). This second triplet has a smaller $D$ coupling and larger $D/E$ ratio, indicating a greater electron-hole separation and a more oblong spin distribution than the $^3$LE. Together with the polarisation pattern,[28] this suggests it arises from a $^3$CT, with hole on Phx and electron on NAI.

Turning to the TEMPO-derivatives, we observe intense, and complex spin-polarised lineshapes in all three compounds. They share the following features: a narrow emissive line is present near $g = 2$; the most intense turning points are narrower, and the full width of the spectrum is broader than the matching Cy-reference compounds. Analogously to the radical-free compounds, the spectrum can be simulated with 1 quartet state for Cz-TEMPO (Fig. 3d) and DMAC-TEMPO (Fig. 3e), but requires 2 for Phx-TEMPO (Fig. 3f). Simulations reveal that these arise from quartet states in the intermediate exchange coupling regime between the triplet and radical. This permits us to directly determine $J_{TR}$, which is near 0.7 GHz for all quartet states detected. Despite this value of $J$, already the X-band spectra are symmetric, since level crossings occur below 100 mT. The optimal zfs values are very similar to those found in respective reference Cy-derivatives, suggesting weak triplet-radical dipolar coupling. We confirm our spectral assignments by performing trESR at Q-band on Cz-TEMPO (Fig. S19), and by explicitly calculating the spectral lineshape for a range of $J_{TR}$ values (Fig. S20). We point out that obtaining the spin Hamiltonian parameters of the parent triplets through studying the Cy-substituted isostructural analogues provides confidence in assigning the complex quartet spectra seen in our TEMPO-derivatives.

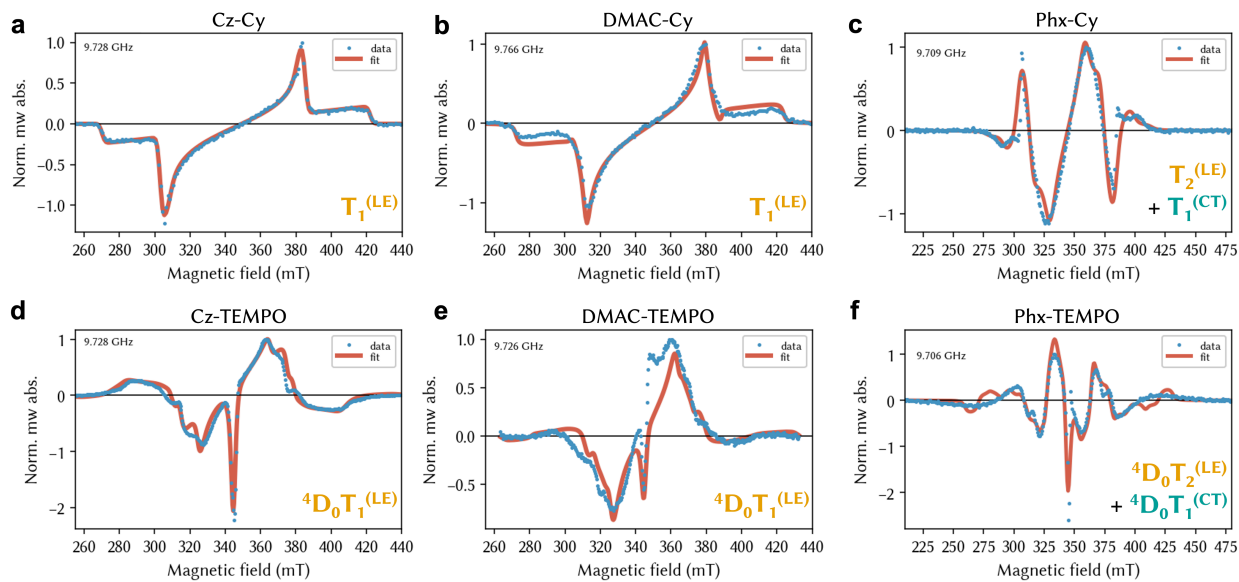

**Fig. 3. Transient ESR.** Prompt (1µs) trESR spectra on frozen 100 µM toluene solutions collected at 80 K at X-band under excitation with 450 nm 1 mJ pulses. The radical-free TADF reference compounds show triplet signatures with **a,** Cz-Cy and **b,** DMAC-Cy spectra displaying an EEEAAA polarisation pattern with near identical turning points. **c,** Phx-Cy spectrum is a linear combination of 2 triplet signatures. The TEMPO-substituted series spectra arise from quartet states in the intermediate exchange coupling regime. **d,** Cz-TEMPO and **e,** DMAC-TEMPO spectra have a similar lineshape, while **f,** Phx-TEMPO spectrum reveals population of 2 quartet species.

| *Material* | $T_n$ | *Weight* | | $\|D_T\|, \|E_T\|$ (MHz) | $P_x:P_y:P_z$ |
|---|---|---|---|---|---|
| Cz-Cy | LE | 100% | | 2133, 53 | 0.53:0.33:0.14 |
| DMAC-Cy | LE | 100% | | 2123, 85 | 0.24:0.72:0.04 |
| Phx-Cy | LE | 88% | | 1920, 120 | 0.28:0.39:0.33 |
| | CT | 12% | | 1515, 236 | 0.15:0.66:0.19 |

| *Material* | $T_n$ | *Weight* | $J_{TR}$ (MHz) | $\|D_T\|, \|E_T\|$ (MHz) | $Q_{+3/2}:Q_{+1/2}:Q_{-1/2}:Q_{-3/2}:D_{+1/2}:D_{-1/2}$ |
|---|---|---|---|---|---|
| Cz-TEMPO | LE | 100% | 700 | 2120, 53 | 0.19:0.09:0.14:0.19:0.21:0.19 |
| DMAC-TEMPO | LE | 100% | 650 | 1923, 130 | 0.21:0.08:0.13:0.22:0.19:0.17 |
| Phx-TEMPO | LE | 71% | 700 | 1920, 120 | 0.23:0.13:0.12:0.25:0.13:0.15 |
| | CT | 29% | 750 | 1515, 236 | 0.07:0.19:0.23:0.05:0.27:0.20 |

**Table 2. Spin Hamiltonian parameters.** Radical-free compounds were simulated as triplets ($S=1$) in the zero-field basis. The TEMPO-derivatives were simulated in the coupled basis ($S_1=1$, $S_2=1/2$).

To explore the potential to manipulate the spin system in our TADF-TEMPO designs, we use pulsed ESR at Q-band (Fig. S21). Cz-TEMPO serves as a long-lived quartet, $^4D_0T_1^{(LE)}$, is analogous to the one found in the DMAC-TEMPO. Transient nutation reveals the entire spectrum is formed of strongly coupled quartet states, as no signals are detected with nutation frequencies corresponding to $S=1$. Quartet spin coherence time is $T_m=1.1$ μs at 80 K in a fully protonated environment. This is comparable to the two key structures of non-luminescent PDI-TEMPO,[11] and luminescent TTM-1Cz-An.[12]

To clarify the energetic landscape of our compounds, we use density functional theory to model the vertical and adiabatic transitions of the closed-shell derivatives (SI Section 6). The adiabatic state diagrams of Cz-Cy and DMAC-Cy are similar, with $^3LE_{NAI}$ remaining the lowest-lying excited state along the curve (Fig. S22). The $^1CT$ and $^3CT$ states of both compounds are almost degenerate in their respective excited state geometries (Tables S3-5). In contrast, Phx-Cy has all three excited states almost iso-energetic at the same geometries, with $^3LE_{NAI}$ above the CT states at the $^1CT$ minimum. This reveals that the energetic ordering of the triplet states depends on molecular conformation for Phx-Cy, which is consistent with the observation of both triplet states contributing to the trESR signals for the Phx compounds.

**Discussion**

We are now able to construct state diagrams for the three TEMPO-derivatives (Fig. 4), based particularly on insights from luminescence spectroscopy and trESR simulations. As radical-triplet exchange is in the intermediate regime for all compounds, the Jabłoński diagrams can be found from matching Cy-compounds, but with each triplet state $T_n$ split into a pair of $^2D_0T_n$ and $^4D_0T_n$ levels separated by a few μeV.

Light absorption in the visible region occurs predominantly along the $^2D_0S_0 \rightarrow {}^2D_0S_1^{(CT)}$ transition on the TADF, due to the low extinction coefficient of TEMPO (Fig. S12). The band gap is determined by the choice of donor unit. Since the position of the $^{2,4}D_0T_n^{(LE)}$ states is the same for all TEMPO-derivatives, the energetic ordering of the LE- and CT-character states is varied through the series. In Cz-TEMPO and DMAC-TEMPO, the lowest energy triplet has LE character, but in Phx-TEMPO the $^{2,4}D_0T_2^{(LE)}$ lie above the CT-character states, leading to both quartets being populated.

DMAC-TEMPO is the prototypical TADF-TEMPO compound, as the energy gap between the $^{2,4}D_0T_n^{(LE)}$ and $^2D_0S_1^{(CT)}$ is of the order of Boltzmann energy at room temperature. This is confirmed by the dominance of delayed emission in films. The absence of $^4D_0T_2^{(CT)}$ features in the ESR indicates that the RISC towards the emissive state is efficient, and does not lead to pooling of excitons in the other quartet state.

For all TEMPO-derivatives, $^2D_0S_1^{(CT)}$ is the only state from which luminescence is detected at room temperature. However, not all observed emission passes through a quartet state. The energy gap between the emissive and the $^4D_0T_1^{(LE)}$ states in Cz-TEMPO leads to the majority of emission occurring before the quartet is formed. In contrast, the well-matched energetics of DMAC-TEMPO allow delayed emission due to RISC from the lowest quartet state to be the dominant luminescence channel.

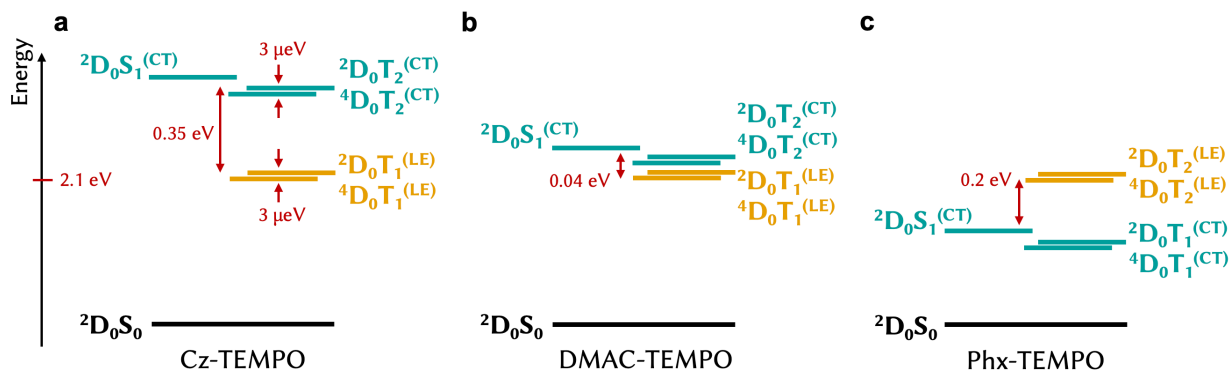

**Fig. 4. Energetics of TEMPO-derivatives.** All three donor units have two triplet states near the emissive singlet, with the position of the CT triplet shifting with the choice of donor and the position of the LE triplet pinned near 2.1 eV (localised on the NAI core). Both triplets split into a pair of trip-doublet and trip-quartet levels separated by a few µeV due to intermediate regime of exchange coupling. **a,** Cz-TEMPO has a high yield of the $^4D_0T_1^{(LE)}$ state but thereafter cannot efficiently emit light from $^2D_0S_1$ due to the substantial ~0.35 eV energy gap. **b,** DMAC-TEMPO is the best TADF-TEMPO structure, as due to thermally accessible energy gaps, RISC from $^4D_0T_1^{(LE)}$ towards $^2D_0S_1$ occurs with an activation of ~0.04 eV. **c,** In Phx-TEMPO the CT states lie below the LE triplet, leading to both $^4D_0T_1^{(CT)}$ and $^4D_0T_2^{(LE)}$ being populated, likely in different molecular conformations.

## Conclusions

This study presents a new design strategy to achieve an emission pathway from a quartet state which does not rely on the presence of a luminescent π-radical. Due to the much greater range of both synthetically accessible non-luminescent radicals and TADF chromophores, compared to the few efficient luminescent radicals, it promises a much greater range of possible structures bearing both electron-rich and electron-deficient functionalities. More broadly, by demonstrating how luminescence can be switched-on in TEMPO-based molecules, it showcases the so far unexplored potential of σ-radicals in optoelectronic research.

While the PLQE can be boosted by engineering the TADF chromophore, improving yield of forward and reverse spin flip processes, and minimising non-radiative losses, the key for optical read-out is maximising not total emission output, but the fraction of RISC from the quartet to the emissive state. In DMAC-TEMPO this is already at a substantial 72% of total emission. The PLQE achieved here is of the same order of magnitude to those in analogous established platforms, such as NV centres in nanodiamonds.[37]

The intermediate exchange regime for the triplet-radical interaction in DMAC-TEMPO may offer additional benefits in applications. As the microwave resonance conditions are very sensitive to the $D/J_{TR}$ ratio in this regime, small perturbations from local magnetic fields may be detectable more easily than in the strong coupling regime in a sensing demonstration.

The TADF-TEMPO design presented here offers a viable pathway to quartet-derived luminescence spanning the entire visible range. More complex molecular structures may then be viable for designing multi-wavelength read-out which could offer new possibilities for interfacing.

# Methods

Transient absorption spectroscopy

Transient absorption experiments were conducted on a setup pumped by a regenerative amplifier (Light Conversion, HARPIA) emitting sub-200 fs pulses centred at 1035 nm at a rate of 10 kHz. The output of the amplifier was optically delayed up to 8 ns by a multi-pass computer-controlled delay stage. This was used to generate a 500 – 1000 nm white light (WL) in a sapphire crystal. For generating a 350 – 500 nm WL, the seed was doubled in a BBO crystal before the sapphire. Wavelength-tunable pump pulses were generated in a Orpheus Neo (Light Conversion) unit. The pump was chopped at 100 Hz to provide on average a 5 kHz repetition rate for pump-on and pump-off measurements. The pump spectrum was filtered using appropriate band-pass filters to remove residual wavelengths, and its polarisation set to magic angle relative to the probe pulses using a Berek rotator. The pump and probe beams were spatially overlapped at the focal point using a beam profiler. The pump and probe diameter at the sample position were on the order of 1000 and 100 µM respectively. After passing through the sample, the probe beam was dispersed with a grating spectrometer (Kymera, Andor Technology) and measured with Si detector arrays.

Transient photoluminescence spectroscopy

Time-resolved PL spectra were collected using an electrically-gated intensified CCD (ICCD) camera (AndoriStar DH740 CCI-010) coupled with an image identifier tube after passing through a calibrated grating spectrometer (Andor SR303i). The spectrometer input slit width was 200 µm. The samples were excited using pump pulses obtained from Orpheus-Lyra (Light Conversion) driven by the same amplifier as the TA setups. The pump repetition rate was reduced to 1 kHz. A suitable long-pass filter was placed directly in front of the spectrometer to avoid the scattered laser signals entering the camera. The kinetics of PL emissions can be obtained by setting the gate delay steps with respect to the excitation pulse. Overlapping time regions were used to compose the decays at several constant gate widths (typically 5 ns, 50 ns, 500 ns, 5 µs and 50 µs). Temperature-dependent measurements were performed using a closed-circuit pressurized helium cryostat (Optistat Dry BL4, Oxford Instruments), compressor (HC-4E2, Sumitomo) and temperature controller (Mercury iTC, Oxford Instruments).

Electron spin resonance

X-band ESR was acquired with a Bruker Biospin E680 or E580 EleXSys spectrometer using a Bruker ER4118-MD5-W1 dielectric $TE01_\delta$ mode resonator (~9.70 GHz) in an Oxford Instruments CF935 cryostat. Q-band ESR employed an EN5107QD2 resonator and a conventional 1.5 T electromagnet like X-band frequencies. The amplifiers for pulsed ESR, Applied Systems Engineering (ASE), had saturated powers of 1.5 kW at X-band and 180 W at Q-band. Temperature was maintained with an ITC-503S temperature controller and a CF-935 helium flow cryostat (both Oxford Instruments). Temperature control was achieved with liquid nitrogen or helium flow and an Oxford Instruments ITC-503s temperature controller.

For laser-induced transient signals, photoexcitation was provided by Ekspla NT230 operating at a repetition rate of 50 Hz. Laser pulse energies used were 0.5 – 1 mJ, pulse lengths of 3 ns transmitted at ca. 40% to the sample via the cryostat, microwave shield and resonator windows. A liquid-crystal depolarizer (DPP-25, ThorLabs) was placed in the laser path for all measurements unless indicated. Triggering of the LASER and ESR spectrometer involved synchronization with a Stanford Research Systems delay generator, DG645. Quadrature mixer detection was used in pulsed and continuous wave detection. Transient cw ESR spectra were simulated using EasySpin.[38,39]

Quantum chemical calculations

All DFT calculations were performed using the ORCA software package.[40] Geometry optimisation of the ground state was performed with CAM-B3LYP functional and the def2-TZVP basis set. Vertical excitations and excited-state geometry optimisations were performed with time-dependent DFT within the Tamm-Dancoff approximation, utilising the root following scheme at the same level of theory as the ground state optimisations. The absence of imaginary frequencies in the calculated Hessians confirmed the stationary nature of converged geometries. Solvent effects were included in all calculations using the polarisable continuum model (PCM) as implemented in Orca, with an $\varepsilon$ of 2.38, corresponding to toluene. Subsequent wavefunction analysis was performed with the Multiwfn package,[41,42] and a combination of homemade scripts.

**Acknowledgments:** We thank Dr Timothy Hele for useful discussions. We also thank Dr Dijana Matak-Vinkovic, Dr Roberto Canales and Asha Boodhun for carrying out the mass spectrometry and Dr Peter Gierth for helpful discussions regarding NMR spectroscopy at the Yusuf Hamied Department of Chemistry, University of Cambridge.

This work was supported by the following funding sources: European Research Council (ERC), European Union's Horizon 2020 research and innovation programme grant agreement no. 101020167 (S.G., P.M., L.M., R.H.F.); John Fell Fund, grant nos.: 0007019, 0010710 (W.K.M.); EPSRC, grant nos: EP/V036408/1 and EP/L011972/1 (CAESR); P.M. has received funding from the European Union's Horizon 2020 research and innovation programme under the Marie Skłodowska-Curie grant agreement No. 891167 and from the Research Council of Finland (No. 363345). D.G.C. has received funding from the Herchel Smith fund and the Royal Society (URF\R1\241806). L.M. thanks the Winton Programme and Harding Distinguished Postgraduate Scholarship for funding. S.G. thanks Emmanuel College, Cambridge for support through a Research Fellowship.

**Author Contributions:** S.G. conceived the project. S.G., D.G.C. and P.M. performed the photophysical measurements. S.G. and W.K.M. performed the ESR measurements. P.M. and D.G.C. synthesized the compounds and performed the chemical characterization. L.M. performed the quantum chemical modelling. A.D.B. carried out the X-ray crystallography and data analysis. S.G. wrote the manuscript with input from all authors. H.B. and R.H.F. provided resources.

**Additional Information:** Supplementary Information is available for this paper. Authors declare that they have no competing interests. The data supporting this study are available at the University of Cambridge Repository [URL to be added].